\documentclass[aoas,MSNbibl,nameyear,seceqn,dvips]{arximspdf}
\usepackage{multirow}
\usepackage{graphicx}

\doi{10.1214/12-AOAS542} 
\volume{6}
\issue{3}
\pubyear{2012}
\firstpage{977}
\lastpage{993}

\makeatletter
\newcommand{\logit}{\operatorname{logit}}
\newcommand{\Perf}{\operatorname{Perf}}
\newcommand{\1}{\mathbf{1}}
\newcommand{\xJ}{\mathcal{J}}
\newcommand{\xM}{\mathcal{M}}
\newcommand{\xR}{\mathbb{R}}
\newcommand{\bC}{\bar{C}}
\makeatother

\begin{document}
\begin{frontmatter}

\title{Classification in postural style}
\runtitle{Classification in postural style}

\begin{aug}
\author{\fnms{Antoine}~\snm{Chambaz}\corref{}\ead[label=e1]{antoine.chambaz@parisdescartes.fr}}
\and
\author{\fnms{Christophe}~\snm{Denis}\ead[label=e2]{christophe.denis@parisdescartes.fr}}
\runauthor{A. Chambaz and C. Denis}
\affiliation{Universit\'e Paris Descartes}
\address{MAP5, UMR CNRS 8145\\
Universit\'e Paris Descartes\\
45 rue des Saints-P\`eres\\
75270 Paris cedex 06\\
France\\
\printead{e1}\\
\phantom{E-mail: }\printead*{e2}}
\end{aug}

\received{\smonth{3} \syear{2011}}
\revised{\smonth{1} \syear{2012}}

%
\begin{abstract}
This article contributes to the search for a notion of \textit{postural
style}, focusing on the issue of classifying subjects in terms of how
they maintain posture. Longer term, the hope is to make it possible to
determine on a case by case basis which sensorial information is prevalent
in postural control, and to improve/adapt protocols for functional
rehabilitation among those who show deficits in maintaining posture,
typically seniors. Here, we specifically tackle the statistical problem
of classifying subjects sampled from a two-class population. Each subject
(enrolled in a cohort of 54 participants) undergoes four experimental
protocols which are designed to evaluate potential deficits in maintaining
posture. These protocols result in four complex trajectories, from which
we can extract four small-dimensional summary measures. Because
undergoing several protocols can be unpleasant, and sometimes painful, we
try to limit the number of protocols needed for the classification.
Therefore, we first rank the protocols by decreasing order of relevance,
then we derive four plug-in classifiers which involve the best
(i.e., more informative), the two best, the three best and all four
protocols. This two-step procedure relies on the cutting-edge
methodologies of targeted maximum likelihood learning (a methodology for
robust and efficient inference) and super-learning (a machine learning
procedure for aggregating various estimation procedures into a single
better estimation procedure). A simulation study is carried out. The
performances of the procedure applied to the real data set (and evaluated
by the leave-one-out rule) go as high as an 87\% rate of correct
classification (47 out of 54 subjects correctly classified), using only
the best protocol.
\end{abstract}

%

\end{frontmatter}

\section{Introduction}
\label{secintro}

This article contributes to the search for a notion of \textit
{postural style},
focusing on the issue of classifying subjects in terms of how they maintain
posture.

Posture is fundamental to all activities, including locomotion and prehension.
Posture is the fruit of a dynamic analysis by the brain of visual,
proprioceptive and vestibular information. Proprioceptive information stems
from the ability to sense the position, location, orientation and
movement of
the body and its parts. Vestibular information roughly relates to the sense
of equilibrium. Every individual develops his/her own preferences according
to his/her sensorimotor experience. Sometimes, a sole kind of information
(usually visual) is processed in all situations. Although this kind of
processing may be efficient for maintaining posture in one's usual
environment, it is likely not adapted to reacting to new or unexpected
situations. Such situations may result in falling, the consequences of
a fall
being particularly bad in seniors. Longer term, the hope is to make it
possible to determine on a case by case basis which sensorial
information is
prevalent in postural control, and to improve/adapt protocols for functional
rehabilitation among those who show deficits in maintaining posture, typically
seniors.

As in earlier studies [\citet{bertrand}, \citet{AntoinePosture} and
references therein],
our approach to characterizing postural control involves the use of a
force-platform. Subjects standing on a force-platform are exposed to
different perturbations, following different experimental protocols (or simply
\textit{protocols} in the sequel). The force-platform records over
time the
center-of-pressure of each foot, that is, ``the position of the global ground
reactions forces that accommodates the sway of the
body'' [\citet{Newelletal97}]. A protocol is divided into three phases:
a first
phase without perturbation, followed by a second phase with perturbation,
followed by a last phase without perturbation. Different kinds of
perturbations are considered. They can be characterized either as
visual, or
proprioceptive, or vestibular, depending on which sensorial system is
perturbed.

We specifically tackle the statistical problem of classifying subjects sampled
from a two-class population. The first class regroups subjects who do
not show
any deficit in postural control. The second class regroups hemiplegic
subjects, who suffer from a proprioceptive deficit. Even though
differentiating two subjects from the two groups is relatively easy by visual
inspection, it is a much more delicate task when relying on some general
baseline covariates and the trajectories provided by a force-platform.
Furthermore, since undergoing several protocols can be unpleasant, and
sometimes painful (some sensitive subjects have to lie down for 15
minutes in
order to recover from dizziness after a series of protocols), we also
try to
limit the number of protocols used for classifying.

Our classification procedure relies on cutting-edge statistical
methodologies. In particular, we propose a nice preliminary ranking of the
four protocols (in view of how much we can learn from them on postural
control) which involves the targeted maximum likelihood
methodology [\citet{vdlaan06}, \citet{TMLEbook}], a statistical
procedure for
robust and
efficient inference The targeted maximum likelihood methodology relies
on the
super-learning procedure, a machine learning methodology for aggregating
various estimation procedures (or simply \textit{estimators}) into a single
better estimation procedure/estimator [\citet{vdlaanSL}, \citet
{TMLEbook}]. In addition
to being a key element of the targeted maximum likelihood ranking of the
protocols, the super-learning procedure plays also a crucial role in the
construction of our classification procedure.

We show that it is possible to achieve an 87\% rate of correct classification
(47 out of 54 subjects correctly classified; the performance is
evaluated by
the leave-one-out rule), using only the more informative protocol. Our
classification procedure is easy to generalize (we actually provide an example
of generalization), so we reasonably hope that even better results are within
reach (especially considering that more data should soon augment our small
data set). The interest of the article goes beyond the specific
application. It nicely illustrates the versatility and power of the targeted
maximum likelihood and super-learning methodologies. It also shows that
retrieving and comparing small-dimensional summary measures from complex
trajectories may be convenient to classify them.

The article is organized as follows. In Section~\ref{secdatadesc} we
describe the data set which is at the core of the study. The classification
procedure is formally presented in Section~\ref{secclassifprocedure}, and
its performances, evaluated by simulations, are discussed in
Section~\ref{secsimstudy}. We report in Section~\ref{secres} the results
obtained by applying the latter classification procedure to the real
data set.
We relegate to the \hyperref[supp]{supplementary file} [\citet{ChaDe12}] a self-contained presentation of the
super-learning procedure as it is used here, and the description of an
estimation procedure/estimator that will play a great role in the
super-learning procedure applied to the construction of our classification
procedure.

\section{Data description}
\label{secdatadesc}

The data set, collected at the Center for the study of sensorimotor functioning
(CESEM, Universit\'e Paris Descartes), is described in
Section~\ref{subsecdatasource}. We motivate the \hyperref[secintro]{Introduction} of a summarized
version of each observed trajectory, and present its construction in
Section~\ref{subsecsummary}.

%
\begin{table}
\caption{Specifics of the four protocols designed to evaluate potential
deficits in postural control. A~protocol is divided into three phases: a
first phase without perturbation of the posture is followed by a second
phase with perturbations, which is followed by a last phase without
perturbation. Different kinds of perturbations are considered. They
can be
characterized either as visual (closing the eyes), or proprioceptive
(muscular stimulation), or vestibular (optokinetic stimulation), depending
on which sensorial information is perturbed}
\label{tabexpDescr}
%
\begin{tabular*}{\textwidth}{@{\extracolsep{\fill}}lccc@{}}
\hline
\textbf{Protocol}& \textbf{1st phase} $\bolds{(0\to15\ \mathrm{s})}$ & \textbf{2nd phase} $\bolds{(15\to50\ \mathrm{s})}$ & \textbf{3rd phase}
$\bolds{(50\to70\ \mathrm{s})}$ \\
\hline
1 & & eyes closed & \\
2 & \multirow{2}{*}{no perturbation} & muscular stimulation & \multirow{2}{*}{no perturbation} \\
3 & & eyes closed & \\
& & muscular stimulation & \\
4 & & optokinetic stimulation & \\
\hline
\end{tabular*}
%
\end{table}

\subsection{Original data set}
\label{subsecdatasource}

Each subject undergoes four protocols that are designed to evaluate potential
deficits in maintaining posture. The specifics of the latter protocols are
presented in Table~\ref{tabexpDescr}. Protocols 1 and 2, respectively, perturb
the processing of visual data and proprioceptive information by the
brain. Protocol~3 cumulates both perturbations. Protocol 4 relies on
perturbing the processing of vestibular information by the brain
through a~visual stimulation.

A total of $n=54$ subjects are enrolled. For each of them, the age, gender,
laterality (the preference that most humans show for one side of their body
over the other), height and weight are collected. Among the 54
subjects, 22
are hemiplegic (due to a cerebrovascular accident), and therefore
suffer from
a proprioceptive deficit in postural control. Initial medical examinations
concluded that the 32 other subjects show no pronounced deficits in postural
control. We will refer to those subjects as \textit{normal subjects}.

\begin{figure}

\includegraphics{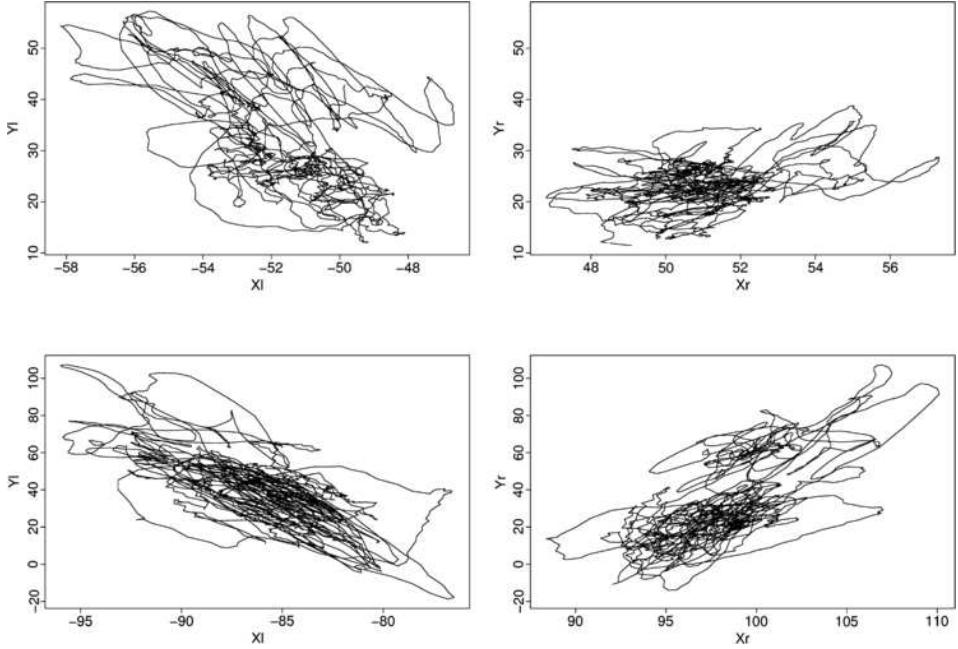}

\caption{Sequences $t \mapsto L_{t}$ (left) and $t \mapsto R_{t}$ (right)
of positions of the center of pressure over $T$ of both feet on the
force-platform, associated with a normal subject (top) and a~hemiplegic
subject (bottom), who undergo the third protocol (see
Table \protect\ref{tabexpDescr}).}\label{fig1}
\vspace*{-3pt}
\end{figure}

For each protocol, the center of pressure of each foot is recorded over
time. Thus, each protocol results in a trajectory $(X_t)_{t \in T} =
(L_t,
R_t)_{t \in T}$, where $L_t = (L_t^1, L_t^2) \in\xR^2$
[resp., $R_t = (R_t^1, R_t^2)$] gives the position of the
center of pressure of the left (resp., right) foot on the
force-platform at time $t$, for each $t$ in $T = \{k \delta\dvtx  1 \leq k
\leq
2800\}$ where the time-step $\delta= 0.025$ seconds (the protocol
lasts 70
seconds). We represent in Figure~\ref{fig1} two such trajectories $(X_{t})_{t
\in T}$ associated with a normal subject and a hemiplegic subject, both
undergoing the third protocol (see Table~\ref{tabexpDescr}). Note that
we do
not take into account the first few seconds of the recording that a generic
subject needs to reach a stationary behavior.

Figure~\ref{fig1} confirms the intuition that the structure of a generic
trajectory $(X_t)_{t \in T}$ is complicated, and that a mere visual inspection
is, at least on this example, of little help for differentiating the normal
and hemiplegic subjects. Although several articles investigate how to model
and use such trajectories directly [\citet{bertrand}, \citet{AntoinePosture}], we rather
choose to rely on a summary measure of $(X_t)_{t \in T}$ instead of
relying on
$(X_t)_{t \in T}$.

\subsection{Constructing a summary measure}
\label{subsecsummary}

\begin{figure}[b]

\includegraphics{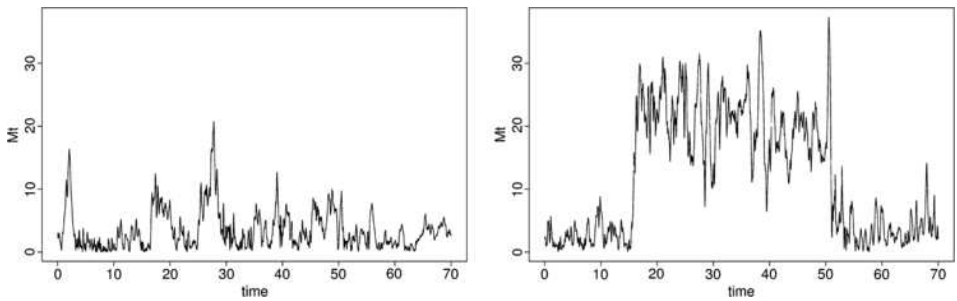}

\caption{Representing the trajectories $t \mapsto
C_{t}$ over $T$ which correspond to two different protocols undergone by
a hemiplegic subject (protocol 1 on the left, protocol 3 on the right).}\label{fig2}
\end{figure}

The summary measure that we construct is actually a summary measure of a
one-dimensional trajectory $(C_{t})_{t \in T}$ that we initially derive
from $(X_{t})_{t \in T}$. First, we introduce the trajectory of barycenters,
$(B_{t})_{t \in T} = (\frac{1}{2} (L_{t} + R_{t}) )_{t \in T}$.
Second, we
evaluate a~reference position $b$ which is defined as the componentwise median
value of $(B_{t})_{t \in T \cap[0, 15]}$ (i.e., the median value
over the
first phase of the protocol). Third, we set $C_t = \| B_t - b \|_2$ for
all $t
\in T$, the Euclidean distance between~$B_{t}$ and the reference
position $b$,
which provides a relevant description of the sway of the body during the
course of the protocol. We plot in Figure~\ref{fig2} two examples of
$(C_{t})_{t \in T}$ corresponding to two different protocols undergone
by a
hemiplegic subject.

\begin{figure}

\includegraphics{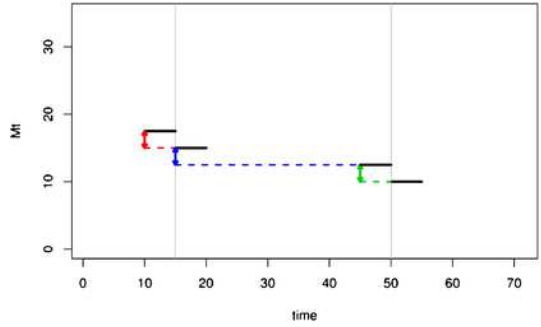}

\caption{Visual representation of the definition of the
finite-dimensional summary measure~$Y$ of $(X_{t})_{t \in T}$. The four
horizontal segments (solid lines) represent, from left to right, the
averages $\bC_{1}^{-}$, $\bC_{1}^{+}$, $\bC_{2}^{-}$, $\bC_{2}^{+}$ of
$(C_{t})_{t \in T}$ over the intervals $[10,15[$, $]15,20]$, $[45,50[$,
$]50,55]$. The three vertical segments (solid lines ending by an arrow)
represent, from top to bottom, the components $Y_{1}$, $Y_{2}$ and $Y_{3}$
of $Y$. Two additional vertical lines indicate the beginning and
ending of
the second phase of the considered protocol.}\label{fig3}
\end{figure}

Because the most informative features can be found at the start and end
of the
second phase, we use the following finite-dimensional\vadjust{\goodbreak} summary measure of
$(X_{t})_{t \in T}$ [through $(C_{t})_{t \in T}$]:
%
%
\begin{equation}
\label{eqsummarymeas}
Y = (\bC_{1}^{+} - \bC_{1}^{-}, \bC_{2}^{-} - \bC_{1}^{+}, \bC
_{2}^{+} -
\bC_{2}^{-}),
\end{equation}
where
\begin{eqnarray*}
\bC_{1}^{-} &=& \frac{\delta}{5} \sum_{t \in T \cap[10,15[} C_{t},
\qquad
\bC_{1}^{+} = \frac{\delta}{5} \sum_{t \in T \cap\,]15,20]} C_{t},\\
\bC_{2}^{-}& =& \frac{\delta}{5} \sum_{t \in T \cap[45,50[} C_{t},\qquad
\bC_{2}^{+} = \frac{\delta}{5} \sum_{t \in T \cap\,]50,55]} C_{t}
\end{eqnarray*}
are the averages of $C_{t}$ computed over the intervals $[10,15[$,
$]15, 20]$,
$[45,50[$ and $]50,55]$ (i.e., over the last/first 5 seconds before/after
the beginning/\break ending of the second phase of the protocol of interest). We
arbitrarily choose this 5-second threshold. Note that $\bC_{2}^{-} -
\bC_{1}^{-} = Y_{2} + Y_{1}$, $\bC_{2}^{+} - \bC_{1}^{-} = Y_{3} + Y_{2}$,
$\bC_{2}^{+} - \bC_{1}^{+} = Y_{1} + Y_{2} + Y_{3}$ are linear
combinations of
the components of $Y$. We refer to Figure~\ref{fig3} for a visual
representation of the definition of the summary measure $Y$.

\section{Classification procedure}
\label{secclassifprocedure}

We describe hereafter our two-step classification procedure. We formally
introduce the statistical framework that we consider in
Section~\ref{subsecstatframe}. The first step of the classification
procedure consists in ranking the protocols from the most to the less
informative with respect to some criterion; see
Section~\ref{subsecstepone}. The second step consists of the classification;
see Section~\ref{subsecsteptwo}.

\subsection{Statistical framework}
\label{subsecstatframe}

The observed data structure $O$ writes as $O = (W,A,Y^{1}, Y^{2},
Y^{3}, Y^{4})$, where\vadjust{\goodbreak}
\begin{itemize}
\item$W \in\xR\times\{0,1\}^{2} \times\xR^{2}$ is the vector of baseline
covariates (corresponding to initial age, gender, laterality, height and
weight, see Section~\ref{subsecdatasource});
\item$A \in\{0,1\}$ indicates the subject's class (with convention
$A=1$ for
hemiplegic subjects and $A=0$ for normal subjects);
\item for each $j \in\{1,2,3,4\}$, $Y^{j} \in\xR^{3}$ is the summary measure
[as defined in~(\ref{eqsummarymeas})] associated with the $j$th protocol.
\end{itemize}

We denote by $P_0$ the true distribution of $O$. Since we do not know much
about $P_{0}$, we simply see it as an element of the nonparametric set
$\xM$
of all possible distributions of $O$.

We need a criterion to rank the four protocols from the most to the less
informative in view of the subject's class. To this end, we introduce the
functional $\Psi\dvtx  \xM\to\xR^{12}$ such that, for any $P \in\xM$,
$\Psi(P)
= (\Psi^{j} (P))_{1 \leq j \leq4}$, where
\[
\Psi^{j} (P) = ( E_{P} \{ E_{P} [Y_{i}^{j} \vert A=1, W]
- E_{P} [Y_{i}^{j} \vert A=0, W ] \} )_{1 \leq i \leq3}.
\]
The component $\Psi_{i}^{j} (P)$ is known in the literature as the variable
importance measure of $A$ on the summary measure $Y_{i}^{j}$
controlling for
$W$ [\citet{TMLEbook}]. Under causal assumptions, it can be interpreted
as the
effect of $A$ on $Y_{i}^{j}$. More generally, we are interested in
$\Psi_{i}^{j} (P_{0})$ because the further it is from zero, the more knowledge
on $A$ we expect to gain from the observation of $W$ and the summary measure
$Y_{i}^{j}$ [i.e., by comparing the averages of $(C_{t})_{t \in T}$
computed over the time intervals corresponding to index~$i$; see
Section~\ref{subsecsummary}]. For instance, say that $\Psi_{1}^{2}
(P_{0}) >
0$: this means that (in $P_{0}$-average) the variation in mean of the mean
postures $\bC_{1}^{-}$ and $\bC_{1}^{+}$ of a~hemiplegic subject computed
before and after the beginning of the muscular perturbation is larger than
that of a normal subject. In words, the postural control of a hemiplegic
subject is more affected by the beginning of the muscular perturbation than
the postural control of a normal subject.

\subsection{Targeted maximum likelihood ranking of the protocols}
\label{subsecstepone}

Our ranking of the four protocols relies on testing the null hypotheses
\[
\mbox{``}\Psi_{i}^{j} (P_0) = 0,\!\mbox{''}\qquad (i,j) \in\{1,2,3\} \times\{1,2,3,4\},
\]
against their two-sided alternatives. Heuristically, rejecting ``$\Psi_{i}^{j}
(P_{0}) = 0$'' tells us that the value of the $i$th coordinate of the summary
measure $Y^{j}$ provides helpful information for the sake of determining
whether $A=0$ or $A=1$.

We consider tests based on the targeted maximum likelihood methodology
[\citet{vdlaan06}, \citet{TMLEbook}]. Because presenting a
self-contained introduction
to the methodology would significantly lengthen the article, we provide below
only a \textit{very succinct} description of it. The targeted maximum
likelihood
methodology relies on the super-learning procedure, a machine\vadjust{\goodbreak}
learning
methodology for aggregating various estimators into a single better
estimator [\citet{vdlaanSL}, \citet{TMLEbook}], based on the
cross-validation principle.
Since super-learning also plays a crucial role in our classification procedure
(see Section~\ref{subsecsteptwo}), and because it is possible to
present a
relatively short self-contained introduction to
the construction of a
super-learner, we propose such an introduction in
the \hyperref[supp]{supplementary file} [\citet{ChaDe12}].

Let $O_{(1)}, \ldots, O_{(n)}$ be $n$ independent copies of $O$. For each
$(i,j) \in\{1,2,3\} \times\{1,2,3,4\}$, we compute the targeted maximum
likelihood estimator (TMLE) $\Psi_{i,n}^{j}$ of $\Psi_{i}^{j}
(P_{0})$ based
on $O_{(1)}, \ldots, O_{(n)}$ and an estimator\vspace*{-1pt} $\sigma_{i,n}^{j}$ of its
asymptotic standard deviation $\sigma_{i}^{j} (P_{0})$. The methodology
applies because $\Psi_{i}^{j}$ is a ``smooth'' parameter. It notably involves
the super-learning of the conditional means $Q_{i}^{j} (P_{0})(A,W) =
E_{P_{0}} (Y_{i}^{j} \vert A,W)$ and of the conditional distribution $g
(P_{0})(A \vert W) = P_{0} (A\vert W)$ (the collection of estimators
aggregated by super-learning is given in the \hyperref[supp]{supplementary file} [\citet{ChaDe12}]). Under some
regularity conditions, the estimator $\Psi_{i,n}^{j}$ of $\Psi_{i}^{j}
(P_{0})$ is consistent when either $Q_{i}^{j} (P_{0})$ or $g (P_{0})$ is
consistently estimated, and it satisfies a central limit theorem. In
addition, if $g(P_{0})$ is consistently estimated by a maximum-likelihood
based estimator, then $\sigma_{i,n}^{j}$ is a conservative estimator of
$\sigma_{i}^{j} (P_{0})$. Thus, we can consider in the sequel the test
statistics $T_{i,n}^{j} = \sqrt{n}\Psi_{i,n}^{j}/\sigma_{i,n}^{j}$
(all $(i,j)
\in\{1,2,3\} \times\{1,2,3,4\}$).

Now, we rank the four protocols by comparing the 3-dimensional vectors
of test
statistics $(T_{1,n}^{j}, T_{2,n}^{j}, T_{3,n}^{j})$ for $1 \leq j \leq4$.
Several criteria for comparing the vectors were considered. They all relied
on the fact that the larger is~$|T_{i,n}^{j}|$ the less likely the null
``$\Psi_{i}^{j} (P_{0}) = 0$'' is true. Since the results were only slightly
affected by the criterion, we focus here on a single one. Thus, we decide
that protocol $j$ is more informative than protocol $j'$ if
\[
\sum_{i = 1}^{3} (T_{i,n}^{j'})^{2} < \sum_{i = 1}^{3} (T_{i,n}^{j})^{2}.
\]
This rule is motivated by the fact that, if $\sigma_{1,n}^{j}$,
$\sigma_{2,n}^{j}$, $\sigma_{3,n}^{j}$ are consistent estimators of
$\sigma_{1}^{j} (P_{0})$, $\sigma_{2}^{j} (P_{0})$, $\sigma_{3}^{j} (P_{0})$,
then $\sum_{i = 1}^{3} (T_{i,n}^{j})^{2}$ asymptotically follows the
$\chi^{2}
(3)$ distribution under $H_{0}^{j}\dvtx ``\Psi^{j} (P_{0}) = 0$.''

By definition of $O$ and by construction of the TMLE procedure, this rule
yields almost surely a final ranking of the four protocols from the
more to
the less informative for the sake of determining whether $A=0$ or $A=1$.

\subsection{Classifying a new subject}
\label{subsecsteptwo}
We now build a classifier $\phi$ for determining whether $A=0$ or
$A=1$ based
on the baseline covariates $W$ and summary measures $(Y^{1}, Y^{2}, Y^{3},
Y^{4})$. To study the influence of the ranking on the classification, we
actually build four different classifiers\vadjust{\goodbreak} $\phi^{1}, \phi^{2}, \phi^{3},
\phi^{4}$ which, respectively, use only the best (more informative) protocol,
the two best, the three best and all four protocols. So $\phi^{j}$ is a
function of $W$ and of~$j$ among the four vectors $Y^{1}, Y^{2}, Y^{3},
Y^{4}$.

Say that $\xJ\subset\{1,2,3,4\}$ has $J$ elements. First, we build an
estimator $h_n^{J} (W,Y^{j},  j \in\xJ)$ of $P_{0}(A = 1 \vert W,
Y^{j}, j \in
\xJ)$ based on $O_{(1)}, \ldots, O_{(n)}$, relying again on the super-learning
methodology (the collection of estimators involved in the
super-learning is
given in the \hyperref[supp]{supplementary file} [\citet{ChaDe12}]). Second, we define
\[
\phi^{J} (W, Y^{j}, j \in\xJ) = \1\bigl\{h_n(W, Y^{j}, j \in\xJ) \geq
{\tfrac{1}{2}}\bigr\}
\]
and decide to classify a new subject with information $(W,Y^{j}, j \in
\xJ)$
into the group of hemiplegic subjects if $\phi^{J}(W,Y^{j}, j \in\xJ
)=1$ or
into the group of normal subjects otherwise.

Thus, the classifier $\phi^{J}$ relies on a plug-in rule, in the sense that
the Bayes decision rule $\1 \{P_{0} (A=1 \vert W, Y^{j}, j \in\xJ)
\geq
{\frac{1}{2}}\}$ is mimicked by the empirical version where one
substitutes an estimator of $P_{0} (A=1 \vert W, Y^{j}, j \in\xJ)$
for the
latter regression function. Such classifiers can converge with fast rates
under a complexity assumption on the regression function and the so-called
margin condition [\citet{AudibertTsybakov07}].

\section{Simulation study}
\label{secsimstudy}

In this section we carry out and report the results of a simulation
study of
the performances of the classification procedure described in
Section~\ref{secclassifprocedure}. The details of the simulation
scheme are
presented in Section~\ref{subsecsimscheme}, and the results are
reported and
evaluated in Section~\ref{subsecloosimul}.

%
\begin{table}[b]
\tablewidth=280pt
\caption{Characterization of the three conditional distributions
$g(P_{0}^{k})$, $k=1,2,3,$ as considered in the simulation scheme}
\label{tabtreatmech}
%
\begin{tabular*}{280pt}{@{\extracolsep{\fill}}l@{}}
\hline
$
\mbox{Scenario 1:} \logit g (P_{0}^{1})(A=1\vert W) =
\displaystyle\frac{W_{1}}{50} + \frac{W_{2}}{50} - \frac{W_{3}}{10} -
\frac{W_{4}}{2000} + W_{5} $\\
$\mbox{Scenario 2:} \logit g (P_{0}^{2}) (A=1\vert W) =
\cos(W_1 + W_5) + \sin(W_1 + W_5)$ \\[2pt]
$\mbox{Scenario 3:} \logit g (P_{0}^{3})(A=1\vert W) =
\lfloor10 \cos(W_1 + W_3) \rfloor$\\[2pt]
$\qquad{}+ \displaystyle\sqrt{5 \cos(W_1 + W_3) - \lfloor5 \cos(W_1 + W_3)
\rfloor} \frac{\pi}{50} \sin\bigl(10 \cos(W_1 + W_3)\bigr)$\\
\hline
\end{tabular*}
\end{table}
%

\subsection{Simulation scheme}
\label{subsecsimscheme}

Instead of simulating $(W, A)$ and the four complex trajectories
$(X_{t}^{1})_{t \in T}$, $(X_{t}^{2})_{t \in T}$, $(X_{t}^{3})_{t \in T}$,
$(X_{t}^{4})_{t \in T}$ associated with four fictitious protocols, we generate
directly $(W, A)$ and the summary measures~$Y^{1}$, $Y^{2}$, $Y^{3}$, $Y^{4}$
that one would have derived from the trajectories $(X_{t}^{1})_{t \in T}$,
$(X_{t}^{2})_{t \in T}$, $(X_{t}^{3})_{t \in T}$, $(X_{t}^{4})_{t \in T}$.
Three different scenarios/probability distributions $P_{0}^{1},
P_{0}^{2}, P_{0}^{3}$ are considered. They only differ from each other with
respect to the conditional distributions $g(P_{0}^{1})$, $g(P_{0}^{2})$,
$g(P_{0}^{3})$ (see Table~\ref{tabtreatmech} for their characterization).

For each $k = 1,2,3$, an observation $O = (W,A,Y^{1}, Y^{2}, Y^{3}, Y^{4})$
drawn\break from~$P_{0}^{k}$ meets the following constraints:
\begin{longlist}[1.]
\item[1.]$W$ is drawn from a slightly perturbed version of the empirical
distribution of $W$ as obtained from the original data set (the same
for all
$k=1,2,3$);
\item[2.] conditionally on $W$, $A$ is drawn from $g (P_{0}^{k}) (\cdot
\vert W)$;
\item[3.] conditionally on $(A,W)$ and for each $(i,j) \in\{1,2,3\} \times
\{1,2,3,4\}$, $Y_{i}^{j}$~is drawn from the Gaussian distribution with mean
$Q_{i}^{j} (A,W)$ (the same for all $k=1,2,3$; see Table \ref
{tabmeans} for
the definition of the conditional means) and common standard deviation
$\sigma\in\{0.5, 1\}$.
\end{longlist}

%
\begin{table}
\caption{Conditional means $Q_{i}^{j} (A,W)$ of $Y_{i}^{j}$ given
$(A,W)$, $(i,j) \in\{1,2,3\} \times\{1,2,3,4\}$, as used in the three
different scenarios of the simulation scheme}\label{tabmeans}
%
\begin{tabular*}{\textwidth}{@{\extracolsep{\fill}}ll@{}}
\hline
\textbf{Fictitious protocol} & \multicolumn{1}{c@{}}{\textbf{Conditional means}} \\
\hline
$j=1$ & $\displaystyle Q_{1}^{1}(A,W)  = 2[A \sin(W_1+W_4) + (1-A) \cos(W_1 +
W_5)]$\\[2pt]
& $\displaystyle Q_{2}^{1} (A,W)  =  3 \biggl[(1-6A) X^{5} - AX^{4} + X^{3} - \biggl(1-\frac{A}{2}\biggr)
X^2 +AX\biggr]$ \\[6pt]
& $\hspace*{143pt}\displaystyle \mbox{where } X = \frac{(1 - 2A)W_{5}}{160} + \frac{A}{4}$\\
& $\displaystyle Q_{3}^{1}(A,W)  =  A \tan(W_{4}) + (1 - A) \tan(W_{5} +
W_{1}W_{2})$\\[3pt]
$j=2$ & $\displaystyle Q_{1}^{2}(A,W)  = \frac{1}{120} [A + W_{1} + W_{2} + W_{3} + W_{5} +
W_{1}W_{2}$\\
& $\hspace*{102pt}\displaystyle {}+ (1-A)W_{5} + W_{2}W_{3}W_{4}]$\\
& $\displaystyle Q_{2}^{2}(A,W)  =  5[A \sin(W_{1}+W_{4}) + (1-A) \cos(W_{1} +
W_{4})]$\\[2pt]
& $\displaystyle Q_{3}^{2}(A,W)  =  \frac{1}{20} \biggl[A\biggl(2W_{1}+\frac{3}{2} W_{3}\biggr) +
(1-A)W_{5}\biggr]$\\[6pt]
$j=3$ & $\displaystyle Q_{1}^{3}(A,W)  =  A \log\biggl(2W_{1}+\frac{3}{2}W_{3}\biggr) + (1-A) \log
(W_{5})$ \\
& $\displaystyle Q_{2}^{3}(A,W)  =  \frac{1}{45} (X+7)(X+2)(X-7)(X-3)$\\
& $\hspace*{165pt}\displaystyle \mbox{where } X = \frac{W_{4} + W_{5}}{145 + AW_{1}}$\\
& $\displaystyle Q_{3}^{3}(A,W)  =  \pi\bigl[ A \sin(X) \bigl(\lfloor2X \rfloor+ \sqrt{2X -
\lfloor2X \rfloor}\bigr)$ \\[2pt]
& $\hspace*{56pt}{}\displaystyle + (1-A) \cos(X) \bigl(\lfloor2X \rfloor+ \sqrt{2X - \lfloor2X
\rfloor}\bigr)\bigr]$ \\[2pt]
& $\hspace*{135pt}\displaystyle \mbox{where } X = \cos(W_{3} + W_{4} + W_{5})$\\[3pt]
$j=4$ & $\displaystyle Q_{1}^{4}(A,W)  =  \frac{1}{100} (2 X^3 +X^2 -X - 1 )$ \\
& $\hspace*{144pt}\displaystyle \mbox{where } X = \frac{AW_2 + W_4+ W_5}{30}$ \\
& $\displaystyle Q_{2}^{4}(A,W)  =  \frac{1}{60} (A + W_{1} + W_{2} + W_{3} +
W_{5})$\\[6pt]
& $\displaystyle Q_{3}^{4}(A,W)  =  \frac{1}{1000} \biggl[ \frac{W_{1}W_{3}W_{4}}{3} +
(1-A)(W_{1} + W_{3}W_{4}) + AW_{2}W_{5} \biggr]$\\
\hline
\end{tabular*}\vspace*{-3pt}
%
\end{table}

\begin{figure}

\includegraphics{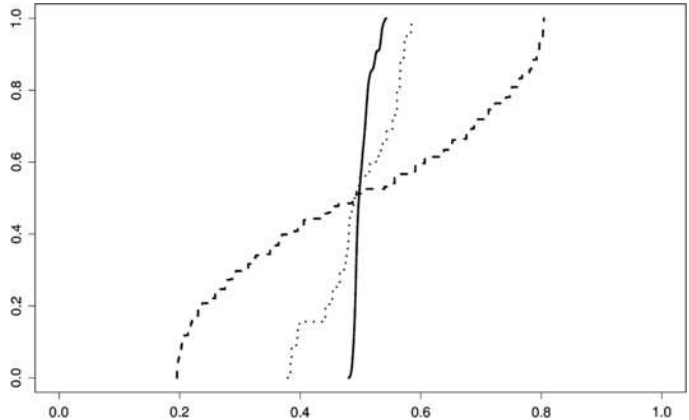}

\caption{Visual representation of the three conditional distributions
considered in the simulation scheme. We plot the empirical cumulative
distribution functions of $\{g^{k} (A=1\vert W_{(\ell)}) \dvtx  \ell= 1,
\ldots, n\}$ for $k=1$ (solid line), $k=2$ (dashed line) and $k=3$
(dotted line), where $W_{(1)}, \ldots, W_{(L)}$ are independent copies
of $W$ drawn from the marginal distribution of $W$ under $P_{0}^{k}$
(which does not depend on $k$), and $L=10^{5}$.}\label{figecdftreatmec}
\end{figure}

Although that may not be clear when looking at Table \ref
{tabtreatmech}, the
difficulty of the classification problem should vary from one scenario
to the
other. When using the first conditional distribution $g(P_{0}^{1})$, the
conditional probability of $A=1$ given $W$ is concentrated around
$\frac{1}{2}$, as seen in Figure~\ref{figecdftreatmec} (solid line), with
$P_{0}^{1} (g (P_{0}^{1}) (1\vert W) \in[0.48,0.54]) \simeq1$. In words,
the covariate provides little information for predicting the class $A$. On
the contrary, estimating $g (P_{0}^{1})$ from the data is easy since
$\logit g
(P_{0}^{1}) (A=1 \vert W)$ is a simple linear function of $W$. The
conditional probabilities of $A=1$ given $W$ under $g (P_{0}^{2})$ and $g
(P_{0}^{3})$ are less concentrated around $\frac{1}{2}$, as seen in
Figure~\ref{figecdftreatmec} (dashed and dotted lines, resp.).
Thus, the covariates may provide valuable information for predicting the
class. But this time, $\logit g (P_{0}^{2})$ and $\logit g (P_{0}^{3})$ are
tricky functions of~$W$.

Likewise, the family of conditional means $Q_{i}^{j} (A,W)$ of $Y_{i}^{j}$
given $(A,W)$ that we use in the simulation scheme is meant to cover a variety
of situations with regard to how difficult it is to estimate each of
them and
how much they tell about the class prediction. Instead of representing the
latter conditional means, we find it more relevant to provide the
reader with
the values (computed by Monte-Carlo simulations) of
\[
S^{j} (P_{0}^{k}) = \sum_{i=1}^{3}\biggl (\frac{\Psi_{i}^{j}
(P_{0}^{k})}{\sigma_{i}^{j} (P_{0}^{k})}\biggr)^{2}
\]
for $(j, k) \in\{1,2,3,4\} \times\{1,2,3\}$ and $\sigma\in\{0.5, 1\}
$; see
Table~\ref{tabtheorranking}. Indeed, $n S^{j} (P_{0}^{k})$ should be
interpreted as a theoretical counterpart to the criterion $\sum_{i=1}^{3}
(T_{i,n}^{j})^{2}$. In particular, we derive from
Table~\ref{tabtheorranking} the theoretical ranking of the protocols: for
every scenario $P_{0}^{k}$ and $\sigma\in\{0.5, 1\}$, the protocols ranked
by decreasing order of informativeness are protocols 3, 2, 1, 4.

%
\begin{table}
\caption{Values of $S^{j} (P_{0}^{k})$ for $(j, k) \in\{1,2,3,4\}
\times\{1,2,3\}$ and $\sigma\in\{0.5, 1\}$. They notably teach us that,
for every scenario $P_{0}^{k}$ and $\sigma\in\{0.5, 1\}$, the protocols
ranked by decreasing order of informativeness are protocols 3, 2, 1, 4}
\label{tabtheorranking}
%
\begin{tabular*}{\textwidth}{@{\extracolsep{\fill}}lcccccc@{}}
\hline
 & \multicolumn{2}{c}{\textbf{Scenario 1}} &
\multicolumn{2}{c}{\textbf{Scenario 2}} & \multicolumn{2}{c@{}}{\textbf{Scenario
3}}\\[-6pt]
& \multicolumn{2}{c}{\hrulefill} &
\multicolumn{2}{c}{\hrulefill} & \multicolumn{2}{c@{}}{\hrulefill}\\
\textbf{Fictitious protocol} & $\bolds{\sigma=0.5}$ & $\bolds{\sigma=1}$ & $\bolds{\sigma=0.5}$ & $\bolds{\sigma=1}$ &
$\bolds{\sigma=0.5}$ & $\bolds{\sigma=1}$ \\
\hline
$j=1$ & 0.14 & 0.04 & 0.11 & 0.03 & 0.14 & 0.04 \\
$j=2$ & 0.86 & 0.37 & 0.74 & 0.31 & 0.85 & 0.37 \\
$j=3$ & 2.94 & 1.12 & 2.49 & 0.93 & 2.90 & 1.11 \\
$j=4$ & 0.06 & 0.01 & 0.04 & 0.01 & 0.06 & 0.01 \\
\hline
\end{tabular*}
%
\end{table}

\subsection{Leave-one-out evaluation of the performances of the classification
procedure}
\label{subsecloosimul}

We rely on the leave-one-out rule to evaluate the performances of the
classification procedure. We acknowledge that they usually result in overly
optimistic error rates. Specifically, we repeat independently $B = 100$ times
the following steps for $k=1,2,3$:
\begin{longlist}[1.]
\item[1.] Draw independently $O_{(1, b)}, \ldots, O_{(n, b)}$ from $P_{0}^{k}$,
with $n=54$; we denote by $A_{(\ell, b)}$ the group membership indicator
associated with $O_{(\ell, b)}$, and by $O_{(\ell, b)}'$ the observed data
structure $O_{(\ell, b)}$ \textit{deprived of} $A_{(\ell, b)}$.
\item[2.] For each $\ell\in\{1, \ldots, n\}$,
\begin{enumerate}[(aaaaaaa)]
\item[(a)] set $\mathcal{S}_{(\ell, b)} = \{O_{(\ell', b)} \dvtx  \ell' \neq
\ell, \ell' \leq n\}$;
\item[(b)] based on $\mathcal{S}_{(\ell, b)}$, rank the protocols (see
Section~\ref{subsecstepone}), then build four classifiers
$\phi^{1}_{(\ell, b)}$, $\phi^{2}_{(\ell, b)}$, $\phi^{3}_{(\ell,
b)}$ and
$\phi^{4}_{(\ell, b)}$ (see Section~\ref{subsecsteptwo}), which,
respectively, use only the best (more informative), the two best, the three
best and all four protocols (thus, $\phi^{J}_{(\ell, b)}$ is a~function of
the covariate $W$ and of $J$ among the four vectors~$Y^{1}$, $Y^{2}, Y^{3},
Y^{4}$);
\item[(c)] classify $O_{(\ell, b)}$ according to the four classifications
$\phi^{1}_{(\ell, b)} (O_{(\ell, b)}')$, $\phi^{2}_{(\ell, b)}
(O_{(\ell,
b)}')$, $\phi^{3}_{(\ell, b)} (O_{(\ell, b)}')$, $\phi^{4}_{(\ell, b)}
(O_{(\ell, b)}')$.
\end{enumerate}
\item[3.] Compute $\Perf^{J}_{b} = \frac{1}{n} \sum_{\ell= 1}^{n} \1\{
A_{(\ell,
b)} = \phi^{J}_{(\ell, b)} (O_{(\ell, b)}')\}$ for $J=1,2,3,4$.
\end{longlist}
From these results, we compute for each $J \in\{1,2,3,4\}$ the mean and
standard deviation of the sample $(\Perf^{J}_{1}, \ldots, \Perf
^{J}_{B})$. All
the standard deviations are approximately equal to 5\%. Second, for every
value of $\sigma\in\{0.5, 1\}$, performance $\Perf^{J}$ actually depends
only slightly on $J$ (i.e., on the number of protocols taken into
account in the classification procedure), without any significant difference
for $j=1,2,3,4$. Third, the latter performances all equal approximately
80\%
when $\sigma= 1$, and increase to approximately 90\% when $\sigma= 0.5$.
This increase is the expected illustration of the fact that the larger
is the
variability\vadjust{\goodbreak} of the summary measures, the more difficult is the classification
procedure. On the contrary, it is a little bit surprising that the
conditional distributions $g(P_{0}^{1}), g(P_{0}^{2}), g(P_{0}^{3})$ do not
affect significantly the performances. Anecdotally, the estimated
ranking of
the protocols always coincide with the ranking that we derived from
Table~\ref{tabtheorranking}.

\section{Application to the real data set}
\label{secres}
We present here the results of the classification procedure of
Section~\ref{secclassifprocedure} applied to the real data set. Thus, we
first rank the protocols from the more to the less informative regarding
postural control (see Section~\ref{subsecranking}); then we construct
the four
classifiers and rely on the leave-one-out rule to evaluate their performances
(see Section~\ref{subsecloo}). A~natural extension of the classification
procedure is considered and applied in Section~\ref{subsecfinal}, and yields
significantly better results. We conclude the article with a
discussion; see
Section~\ref{subsecdisc}.

\subsection{Targeted maximum likelihood ranking of the protocols over
the real
data set}
\label{subsecranking}

Hemiplegic subjects are known to be sensitive to muscular stimulations, and
also to tend to compensate for their proprioceptive deficit by
developing a
preference for visual information in order to maintain
posture [\citet{bonan96}]. This suggests that protocols involving muscular
and/or visual stimulations should rank high. What do the data tell us?

We derive and report in Table~\ref{tabranking} the results of the
ranking of
the protocols using the entire data set. Table~\ref{tabranking}
teaches us
that the most informative protocol is protocol~3 (visual \textit{and} muscular
stimulations), and that the three next protocols ranked by decreasing
order of
informativeness are protocols~2 (muscular stimulation), 1~(visual
stimulation) and 4 (optokinetic stimulation). Apparently, protocols 3
and~2
(which have in common that muscular stimulations are involved) are highly
relevant for differentiating normal and hemiplegic subjects based on postural
control data. On the contrary (and perhaps surprisingly, given the
introductory remark), protocols 1 and 4 seem to provide significantly less
information for the same purpose.

%
\begin{table}
\caption{Ranking the four protocols using the entire real data set. We
report the realizations of the criteria $\sum_{i = 1}^{3}
(T_{i,n}^{j})^{2}$ obtained for protocols $j=1,2,3,4$. These values teach
us that the most informative protocol is protocol 3, and that the three
next protocols ranked by decreasing order of informativeness are
protocols 2, 1 and 4}
\label{tabranking}
%
\begin{tabular*}{\textwidth}{@{\extracolsep{\fill}}lcccc@{}}
\hline
\textbf{Protocol} & $\bolds{j = 3}$ & $\bolds{j = 2}$ & $\bolds{j = 1}$ & $\bolds{j = 4}$ \\
\hline
Criterion $\sum_{i = 1}^{3} (T_{i,n}^{j})^{2}$ & 75.51 & 33.13 & 6.80 &
5.53\\
\hline
\end{tabular*}
%
\end{table}

\subsection{Classification procedures applied to the real data set}
\label{subsecloo}

To evaluate the performances of the classification procedure applied to the
real data set, we carry out steps 2a, 2b, 2c from the leave-one-out rule
described in Section~\ref{subsecloosimul},\vadjust{\goodbreak} where we substitute the real
data set $O_{(1)}, \ldots, O_{(n)}$ for the simulated one. We actually
do it
\textit{twice}. The first time, the super-learning methodology
involves a large
collection of estimators; the second time, we justify resorting to a smaller
collection (see the \hyperref[supp]{supplementary file} [\citet{ChaDe12}]). We report the results in
Table~\ref{tabperf}, where the second and third rows, respectively, correspond
to the first (larger collection) and second (smaller collection) rounds of
performance evaluation.

Consider first the performances of the classification procedure relying
on the
larger collection. The proportion of subjects correctly classified (evaluated
by the leave-one-out rule) equals only 70\% (38 out of the 54 subjects are
correctly classified) when the sole most informative protocol (i.e.,
protocol~3) is exploited. This rate jumps to 80\% (43 out of 54
subjects are
correctly classified) when the two most informative protocols (i.e.,
protocols 3 and 2) are exploited. Including one or two of the remaining
protocols decreases the performances.

%
\begin{table}
\caption{Leave-one-out performances $\Perf^{J}$ of the classification
procedure using the real data set. Performance $\Perf^J$ corresponds
to the
classifier based on $J$ among the four vectors~$Y^{1}$, $Y^{2}, Y^{3},
Y^{4}$ (those associated with the $J$ more informative protocols) and
either using all estimators (second row) or only two of them
(third row) in the super-learner (see Appendix A in the \protect\hyperref[supp]{supplementary file} [\citet{ChaDe12}])}
\label{tabperf}
%
\begin{tabular*}{\textwidth}{@{\extracolsep{\fill}}lcccc@{}}
\hline
& $\bolds{J= 1}$ & $\bolds{J =2}$ & $\bolds{J = 3}$ & $\bolds{J = 4}$ \\
\hline
$\Perf^{J}$ (all est.) & 0.70 (38/54) & 0.80 (43/54) & 0.74 (40/54)
& 0.78 (42/54) \\
$\Perf^{J}$ (two est.) & 0.74 (40/54) & 0.81 (44/54) & 0.78 (42/54)
& 0.85 (46/54) \\
\hline
\end{tabular*}
%
\end{table}

The theoretical properties of the super-learning procedure are asymptotic,
that is, valid when the sample size $n$ is large, which is not
the case in
this study. Even though this is contradictory to the philosophy of the
super-learning methodology, it is tempting to reduce the number of estimators
involved in the super-learning. We therefore keep only two of them, and run
again steps 2a, 2b, 2c from the leave-one-out rule described in
Section~\ref{subsecloosimul}, where we substitute the real data set $O_{(1)},
\ldots, O_{(n)}$ for the simulated one. Results are reported in
Table~\ref{tabperf} (third row). We obtain better performances: for each
value of $J$ (i.e., each number of protocols taken into account
in the
classification procedure), the second classifier outperforms the first one.
The best performance is achieved when all four protocols are used,
yielding a
rate of correct classification equal to 85\% (46 out of the 54 subjects are
correctly classified). This is encouraging, notably because one can reasonably
expect that performances will be improved on when a larger cohort is
available.

Yet, this is not the end of the story. We have built a general methodology
that can be easily extended, for instance, by enriching the small-dimensional
summary measure derived from each complex trajectory. We explore the effects
of such an extension in the next section.

\subsection{Extension}
\label{subsecfinal}

Thus, we enrich the small-dimensional summary measure initially defined in
Section~\ref{subsecsummary}. Since it mainly involves \textit{distances}
from a
reference point, the most natural extension is to add information pertaining
to \mbox{\textit{orientation}}. Relying on polar coordinates of the trajectory
$(B_{t})_{t \in T}$ poses some technical issues. Instead, we propose
to fit
simple linear models $y(B_{t}) = v x(B_{t}) + u$ [where $x(B_{t})$ and
$y(B_{t})$ are the abscisse and ordinate of $B_{t}$] based on the data sets
$\{B_{t} \dvtx  t \in T \cap[10, 15[\}$, $\{B_{t} \dvtx  t \in T \cap[15, 20[\}$,
$\{B_{t} \dvtx  t \in T \cap[20, 45[\}$, $\{B_{t} \dvtx  t \in T \cap[45,
50[\}$ and
$\{B_{t} \dvtx  t \in T \cap[50, 55[\}$, and to use the \textit{slope
estimates} as
summary measures of an average orientation over each time interval. The
observed data structure and parameter of interest still write as $O=(W, A,
Y^{1}, Y^{2}, Y^{3}, Y^{4})$ and $\Psi(P) = (\Psi^{j} (P))_{1 \leq j
\leq
4}$, but $Y^{j}$ and $\Psi^{j} (P)$ now belong to $\xR^{8}$ (and not
$\xR^{3}$ anymore). The ranking of the protocols now relies on the criterion
$\sum_{i=1}^{8} (T_{i,n}^{j})^{2}$, whose definition straightforwardly extends
that of the criterion introduced in Section~\ref{subsecstepone}. The values
of the criteria are reported in Table~\ref{tabrankingextended}. The ranking
of protocols remains unchanged, but the discrepancies between the
values for
protocol 2, on one hand, and for protocols 1 and 4, on the other hand, are
smaller.

%
\begin{table}
\caption{Ranking the four protocols using the entire real data set
\textit{and the extended small-dimensional summary measure} of the complex
trajectories. We report the realizations of the criteria $\sum_{i = 1}^{8}
(T_{i,n}^{j})^{2}$ obtained for protocols $j=1,2,3,4$. The ranking is the
same as that derived from Table \protect\ref{tabranking}}
\label{tabrankingextended}
%
\begin{tabular*}{\textwidth}{@{\extracolsep{\fill}}lcccc@{}}
\hline
\textbf{Protocol} & $\bolds{j = 3}$ & $\bolds{j = 2}$ & $\bolds{j = 1}$ & $\bolds{j = 4}$ \\
\hline
Criterion $\sum_{i = 1}^{8} (T_{i,n}^{j})^{2}$ & 83.64 & 43.61 & 14.92 &
12.60\\
\hline
\end{tabular*}
%
\end{table}

We finally apply once again steps 2a, 2b, 2c from the leave-one-out rule
described in Section~\ref{subsecloosimul}, where we substitute the real
data set $O_{(1)}, \ldots, O_{(n)}$ for the simulated one, and use
either all
estimators or only two of them in the super-learner. The results are reported
in Table~\ref{tabperfextended}.

%
\begin{table}[b]
\caption{Leave-one-out performances $\Perf^{J}$ of the classification
procedure using the real data set \textit{and the extended small-dimensional
summary measure} of the complex trajectories. Performance $\Perf^{J}$
corresponds to the classifier based on $J$ among the four vectors~$Y^{1}$,
$Y^{2}, Y^{3}, Y^{4}$ (those associated with the $J$ more informative
protocols) and either using all estimators (second row) or only
two of them (third row) in the super-learner (see Appendix A
in the \protect\hyperref[supp]{supplementary file} [\protect\citet{ChaDe12}])}
\label{tabperfextended}
%
\begin{tabular*}{\textwidth}{@{\extracolsep{\fill}}lcccc@{}}
\hline
& $\bolds{J= 1}$ & $\bolds{J =2}$ & $\bolds{J = 3}$ & $\bolds{J = 4}$ \\
\hline
$\Perf^{J}$ (all est.) & 0.82 (44/54) & 0.80 (43/54) & 0.80 (43/54)
& 0.78 (42/54) \\
$\Perf^{J}$ (two est.) & 0.87 (47/54) & 0.85 (46/54) & 0.80 (43/54)
& 0.82 (44/54) \\
\hline
\end{tabular*}
%
\end{table}

When we include all estimators in the super-learner, the classification
procedure that relies on the extended small-dimensional summary measure
of the
complex trajectories outperforms the classification procedure that
relies on
the initial summary measure, for every value of $J$ (i.e., each number
of protocols taken into account in the classification procedure). The
performances are even better when we only include two estimators. Remarkably,
the best performance is achieved using only the most informative protocol,
with a proportion of subjects correctly classified (evaluated by the
leave-one-out rule) equal to 87\% (47 out of the 54 subjects are correctly
classified).\looseness=1

\subsection{Discussion}
\label{subsecdisc}
We conducted a brief simulation study to evaluate the performances of the
classification procedure. With its three different scenarios [i.e.,
three conditional distribution $g(P_{0}^{k})$] and four trajectories
(i.e., twelve conditional means $Q_{i}^{j}$), the simulation scheme is far
from comprehensive. Rather than extending the simulation study, we discuss
here what additional scenarios would need to be considered before
applying the
procedure more generally. In the same spirit as in
Section~\ref{secsimstudy}, one should consider the following:
\begin{itemize}
\item other conditional distributions $g(P_{0}^{k})$,
$|g(P_{0}^{k}(A=1|W) -
1/2|$ being close to~0 with high probability ($W$ strong predictor of $A$)
or low probability ($W$ weak predictor of $A$);
\item other conditional means $Q_{i}^{j}$, $(i,j) \in\{1,2,3\} \times
\{1,2,3,4\}$, and standard deviation $\sigma$, $\{S^{j} (P_{0}^{k})\dvtx j =
2,3,4\}$ having one, two, three or four well-separated values.
\end{itemize}
A straightforward generalization would consist in allowing the standard
deviation of $Y_{i}^{j}$ to depend on $(i,j)$. Furthermore, another approach
to simulating could be considered, where the trajectories
$(X_{t}^{1})_{t \in
T}$, $(X_{t}^{2})_{t \in T}$, $(X_{t}^{3})_{t \in T}$,
$(X_{t}^{4})_{t \in
T}$ would be obtained as realizations of stochastic processes
satisfying a
variety of piecewise stochastic differential equations (SDEs). For instance,
the same SDE could be used to simulate the trajectory during the first and
third phases ($0\to15$~s and $50\to70$~s, without perturbations), and another
SDE could be used to simulate during the second phase ($15\to50$~s, with
perturbations). On top of that, the breaking points could be drawn randomly
from two symmetric distributions centered at 15~s and 50~s.

This alternative approach to simulating arose while we were trying to quantify
in some way how much information is lost when one substitutes a~summary
measure for the original trajectory for the purpose of classifying.
Ultimately such a quantification could permit to elaborate new summary
measures with minimal information loss. We did not obtain a satisfactory
answer to this very difficult question. However, we identified important
information that can be derived from the original trajectory, such as mean
orientation, as used in Section~\ref{subsecfinal}, and empirical breaking
points, as evoked for the sake of simulating in the previous paragraph, and
used for the sake of classifying by \citet{Denis11}.

\section*{Acknowledgments}
The authors thank I. Bonan (Service de M\'edecine Physique et de
r\'eadaptation, CHU Rennes) and P.-P. Vidal (CESEM, Universit\'e Paris
Descartes) for introducing them to this interesting problem and
providing the
data set. They also thank warmly A. Samson (MAP5, Universit\'e Paris
Descartes) for several fruitful discussions, and the Editor for suggesting
improvements.

\begin{supplement}\label{supp}
\sname{Supplementary file}
\stitle{Supplement to ``Classification in postural style''}
\slink[doi]{10.1214/12-AOAS542SUPP} 
\slink[url]{http://lib.stat.cmu.edu/aoas/542/supplement.pdf}
\sdatatype{.pdf}
\sdescription{We gather in this Supplementary file a~short and
self-contained description of the construction of a~super-learner, as well
as the estimation procedures that we choose to involve for the sake of
classifying subjects in postural style. One of those estimation
procedures, a variant of the top-scoring pairs classification procedure,
is specifically presented.}
\end{supplement}

%

\printaddresses

\end{document}